\documentclass[11pt]{article}
\usepackage[dvipdfmx]{graphicx}
\usepackage[dvipdfmx]{color}
\usepackage{color}

\usepackage{amsmath,amssymb}

\usepackage{subcaption}
\begin{document}

\title{Understanding the effect of the base oil on the physical adsorption process of organic additives using molecular dynamics}
\author{Masakazu Konishi and Hitoshi Washizu
\footnote{Graduate School of Simulation Studies, University of Hyogo, 7-1-28 Minatojima-minamimachi, Chuo-ku, Kobe, Hyogo 650-0047, Japan, 
Elements Strategy Initiative for Catalysts and Batteries (ESICB), Kyoto University, 1-30 Goryo-Ohara, Nishikyo-ku, Kyoto 615-8245, Japan, h@washizu.org}}
\maketitle
\begin{abstract}
Organic friction modifiers (OFMs) are widely added to oil to reduce the boundary friction in many kinds of lubricants such as vehicle engine oils. At the contact area in machine elements, the OFMs form a self-assembled organic monolayer. Although the friction properties of the monolayer are widely studied on a molecular level, the formation process is not well-known. In this study, the initial adsorbing process of additive molecules in explicit base oil molecules are calculated using molecular dynamics. The adsorption time depends on the structure of the base oils. Another effect of the base oil other than "chain matching" is found.
\end{abstract}  

\section{Introduction}

In machine elements, 
there are many interfaces that perform relative motions,
and it is necessary to control the friction and wear generated
between the two surfaces.
The control of friction and wear leads to the suppression of energy loss,
vibration,
improvement of function, performance, and reliability
at the entire mechanical system.
In order to inhibit the contact between the two surfaces,
hydrodynamic lubrication by oil may be the best solution.
There are, however, many systems that cannot maintain the
hydrodynamic lubrication film due to the sliding condition.
A boundary lubrication film is the candidate to decrease the
friction in such solid contacts.
Organic friction modifiers (OFMs) or
oiliness agents are some of the most popular additives which 
produce a boundary lubrication film~\cite{Bowden:1964}.

OFM additive molecules are composed of 
alkyl chains and polar groups on the end. 
Fatty acids are the 
first used OFMs~\cite{Spikes:2015}.
The polar group adsorbs on the metal surface and
the molecules form a self-assembled monolayer (SAM).
The mechanical and physiochemical stability of the
monolayer is important to protect the metal surface
and to reduce friction. 

In the long history of the friction modifiers,
OFMs are the most common,
which have been used since the 1920s, and widely studied~\cite{Spikes:2015}.
The most important idea is the formation of an adsorbed
self-organized layer by Bowden and Tabor~\cite{Bowden:1964}.
After this model, the concept of "chain matching'' by
Askwith et al.~\cite{Askwith:1966} is the important
finding in order to consider the effect of
the base oil (solvent).
If the chain length of the fatty acid additive is same as
the chain length of the base oil, the system shows
a lower friction than other base oils. 
Although this idea is supported by experiments, such as the
experiments 
focused on nucleation and crystal growth~\cite{Hirano:1987},
the precise mechanism of low friction is still
an open question~\cite{Cameron:1997}.

Molecular dynamics (MD) simulations are used to
understand the friction phenomena of SAMs,
since the MD is the most useful tool to
understand the dynamics of SAMs and
comparable with the experimental data
obtained by
a surface force apparatus (SFA) or
atom force microscope (AFM)~\cite{Spikes:2015}.
The MD simulation demonstrated that
the friction behavior of the SAMs obtained by
AFM experiments is well described by
the simulation between the
hard AFM tip and the soft SAMs~\cite{koike, koike2, ohzono, cummings}.
This was one of the most studied friction systems
using MD during the early stage of the nano tribology studies. 
Recent nonequilibrium molecular dynamics (NEMD) 
studies show moer detailed structure and friction behavior of OFMs
~\cite{Doig, Ewen2016a, Ewen2018},
and NEMD of OFM reverse micelles adsorption behaviour under shear
~\cite{Bradley-Shaw2016, Bradley-Shaw2016}.
Tight-binding quantum chemistry calculations
~\cite{Loehle2014, Loehle2015} and
density functional theory claculations
~\cite{Gattinoni}.
 have also been used to study OFM adsorption 
~\cite{Loehle2014, Loehle2015, Gattinoni}.
These methods are required to accurately reproduce strong chemisorption observed in XPS experiments
~\cite{Loehle2015},
but are computationally expensive and limited in terms of accessible system size and timescales.

The formation process of the monolayers, on the other hand,
is not yet well known. 
In experiments, adsportion process is observed by
polarized neutron reflectometry
~\cite{Campana2011} or AFM~\cite{Campana2015},
however, the observation of the initial process is 
difficult.

The major part of the tribological use of the OFMs
used in the solution of the base oil is more than 90~\%,
and a small amount of additives.
In the engine oil of the automobile, the percentage of the
OFMs are only a few percent. 
It is known that only 0.5~\% OFMs makes
the friction lowering effect significantly~\cite{Spikes:2015}.
When simulating the solution including 4~\% additives by MD,
about 96~\% of the calculation cost is devoted to the
calculation of the fluctuation of the base oil molecules.
The formation process of the SAMs using a coarse-grained
MD simulation is proposed~\cite{miura}.
Both the adsorption process of the additives and the
formation of SAM by aggregation of the adsorbed molecules
are simulated. 
In the simulation, however, the solvent molecules are
described as a single sphere molecule using 
the Lennard-Jones attractive or soft-core repulsive interactions.
The effect of the structure of solvent
molecules can not be discussed by the sphere model in 
the coarse-grained simulation.
In order to directly know the solvent effect, such as chain matching,
the all-atom MD simulation is more appropriate. 
All-atom force-fields have also important to accurately model OFM film strucure and friction
~\cite{Ewen2016b}.

In this study, the all-atom MD simulation is used to solve the
solvent (base oil) effect on the formation process of the
OFM film. Since the calculation cost of the base oil is
high, only the beginning process of the adsorprtion is
discussed here. However, an interesting effect,
i.e., the adsorption time varies with the structure
of the layer of the base oil molecules is found. 

\section{Simulation Methods}

\begin{figure}[htbp]
 \begin{center}
  \includegraphics{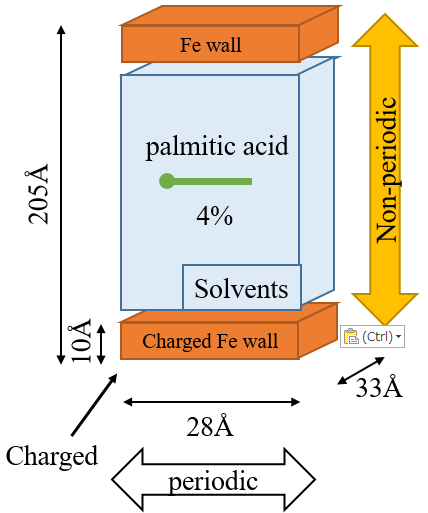}
 \end{center}
 \caption{Schematic picture of the simulation model used in
 the study. Boundary conditions are also shown.}
 \label{fig:sim_model}
\end{figure}

\begin{figure}[htbp]
 \begin{minipage}{0.24\hsize}
   \begin{center}
    \includegraphics{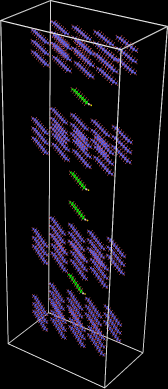}
   \end{center}
    \subcaption{Initial distribution of the bulk solution.}
    \label{fig:state1}
 \end{minipage}
 \begin{minipage}{0.24\hsize}
   \begin{center}
    \includegraphics{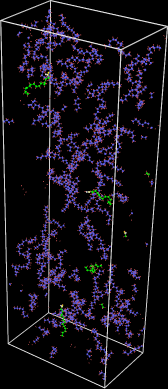}
   \end{center}
   \subcaption{Annealing process of the bulk solution.}
    \label{fig:state2}
 \end{minipage}
 \begin{minipage}{0.24\hsize}
   \begin{center}
   \includegraphics{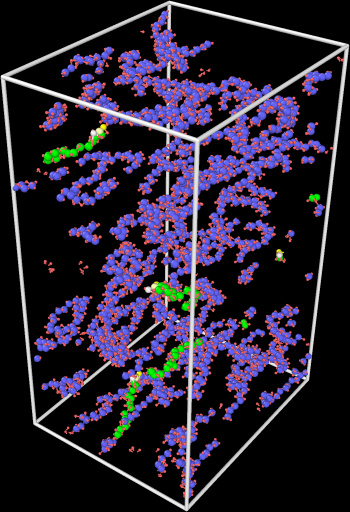}
  \end{center}
   \subcaption{Pressing process of the bulk solution.}
   \label{fig:state3}
 \end{minipage}
 \begin{minipage}{0.24\hsize}
  \begin{center}
   \includegraphics{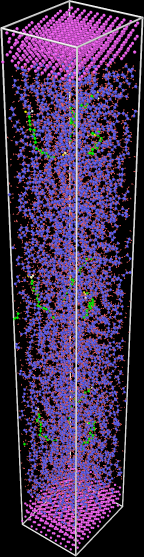}
  \end{center}
  \subcaption{Initial distribution for the adsorption simulation.}
  \label{fig:state4}
 \end{minipage}
 \caption{Schematic snapshots of preparation of initial distribution
   of the solution in a confined geometry.
   Base oil is n-hexadecane and oiliness additive molecule is palmitic acid.
   The color of the atoms are as follows: pink; ferrous, blue; carbon
   atoms in the base oil molecules,
   red; hydrogen, green; carbon atoms in additive molecules,
   wheat; carbon atom of a carboxylic acid group,
   yellow and white; oxigen atom of a carboxylic acid group.
   }
  \label{fig:model_make}
\end{figure}

All-atom MD simulations of the adsorption process of additives (OFMs) are done
in the following manner.
Schematic picture of the simulation model is shown in Fig.~\ref{fig:sim_model},
and step by step procedure for preparing the model solution is shown in 
Fig.~\ref{fig:model_make}.

The solution with the hydrocarbon base oil and additives is confined
by two solid layers. The top solid layer is a neutral Fe wall and
the bottom solid layer is a charged Fe wall (Fig.~\ref{fig:sim_model}). 
The Fe walls are modeled as a universal model of solid materials.
Then the charge on the bottom solid layer is put, then the adsorption
process was simulated. The precise conditions are described as follows.

First, three types of solutions of 
the base oil with additives under standard conditions is prepared.
The base oil (solvent) of n-hexadecane, 2, 4-dimethyltetradecane, 
and 3, 5-dimethydodecane with the additive (solute) of palmitic acid
was chosen as the linear and branched base oils with 
the solute of the same number of carbons.
The latter branched base oils are taken from the structure of 
the Olefin Copolymer, or Poly $\alpha$ olefin.
In the MD simulation,
the organic molecules are dynamically treated using 
the Dreiding force field~\cite{dreiding}. 
This all-atom force field include bonds, angles, dihiderals (torsions) and improper torsions 
for the intra-molecular forces, and Lennard-Jones and Coulomb for the inter-molecular forces.
The partial atomic charges on the organic molecules are determined using 
the MOPAC6~\cite{mopac}  semi-empirical molecular orbital calculation 
with the Hamiltonian:AM1.

A set of organic molecules is first arranged in a lattice configuration. 
96 base oil molecules and 4 solute molecules are arranged
in order to set the concentration of the additive at 4~\%
(Fig.~\ref{fig:model_make}(a)), 
which is the commonly used condition such as in engine oil for automobiles.
The size of the simulation box was 27.31~\AA~along the $x$ axis, 32.77~\AA~along 
the $y$ axis and an arbitrary length for the $z$ axis, since the system 
is then pressed.
The molecules are
then moved by the MD simulation under the periodic boundary condition in
the $x,y$ directions, and reflective wall in the $z$ direction,
for 500~ps at the constant temperature of 1,000~K in order to anneal the system
(Fig.~\ref{fig:model_make}(b)).

The MD simulation is then done for 10~ns in order to
cool the system to 300~K.
The temperature is controlled using the Nose-Hoover
thermostat~\cite{nvt, hoover, nvt_text_ref}. 
Then the system is pressed to the $z$ direction so that the
density of the base oil is the same as the experimental value~\cite{npt}.
The solution in a thermal equilibrium is obtained (Fig.~\ref{fig:model_make}(c)). 
Next, 3 sets of thermal equilibrium fluids
are arranged in the $z$ direction between the two solid plates,
(Fig.~\ref{fig:model_make}(d)), in order to form a semi-bulk region in the center of the
oil film. The final thickness of the solution in the $z$ direction is 194.19~\AA.
On the $x$ and $y$ axes, periodic boundary conditions are adopted. 
On the $z$ axe, non-periodic boundary conditions are adopted.

All the organic molecules are connected to a Nose-Hoover thermostat. 
The long-range Coulombic interactions are calculated by the Multi-Summation Method~\cite{msm}. 
This method enables to simulate a non-periodic system.
The equations of motion were integrated using a velocity-Verlet method~\cite{verlet}
with a time step of 2.0~fs. 
This time step is long for all-atom simulation.
During the simulation, a drift of conserved quantity such as temperature
are not found. Since the phenomena is governed by long-range
coulomb interaction between the surface and the functional group,
and the dynamics of not the hydrogen-carbon but carbon-carbon bonds are critical, 
we think this phenomena will reproduce at least qualitatively in
shorter time step.

Each system was calculated five times, 
while changing the initial speed.
The adsorption dynamics changes drastically by initial speed.

The universal solid plate (wall) to analyze the adsorption process is made by
the following procedure. We propose this in order to treat both charged oxidized metal
surfaces and the neutral metal surface with the least difference, i.e., the existance
of a charge on the surface. 
If the chemical reality of the surface, 
including the chemical reactions is in interest, 
a more realistic model 
such as copper and copper oxide surface
treated by reactive force field should be used~\cite{Washizu:2018}.
We are now interested in the physical nature of the interface and this model
succeeded in reproducing the formation of the elastohydrodynamic lubrication 
oil film~\cite{washizu_lubri, Washizu:2014, Washizu:2017}.
The parameter for the solid is taken from an alpha-ferrous crystal,
a solid atom layer with a lattice of 
11 $\times$ 7 $\times$ 5 atoms
in the $x,y$ and $z$ directions, respectively, 
and the lattice parameter is set to 2.87~\AA. 
The fluid-solid interface is the (110) surface, 
and the vibrations of each solid atom are suppressed.
These atoms are not connected to the Nose-Hoover thermostat.

The parameter for the solid atoms is then set to 0.2853 Kcal/mol~\cite{washizu_lubri}. 
This parameter is taken from previous study~\cite{Tamura} and it is known that
the macroscopic slip between solid atoms and fluid molecules under sliding motion
are supressed~\cite{washizu_lubri}.

\begin{figure}[htbp]
 \begin{center}
  \includegraphics{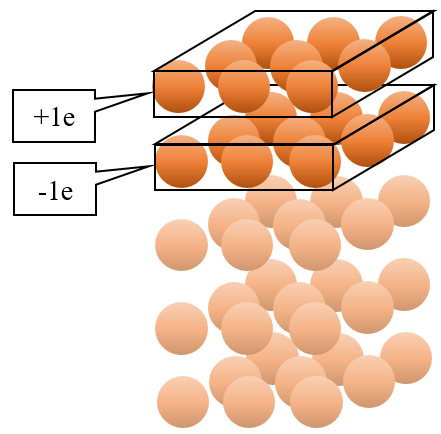}
 \end{center}
 \caption{Schematic picture of the solid wall model.
 The uppermost layer is charged +1e, and the
 second layer is charged -1e.}
 \label{fig:solid_wall_model}
\end{figure}

We then arranged only the bottom solid wall which is charged. 
The charge distribution on the surface of the solid wall is shown in 
Fig.~\ref{fig:solid_wall_model}. 
The outermost layer is charged at +1e,
and the second layer is charged at -1e in order to make system 
electrical neutral.

\section{Results and Discussion}

\begin{figure}[htbp]
 \begin{minipage}{0.5\hsize}
  \begin{center}
    \includegraphics{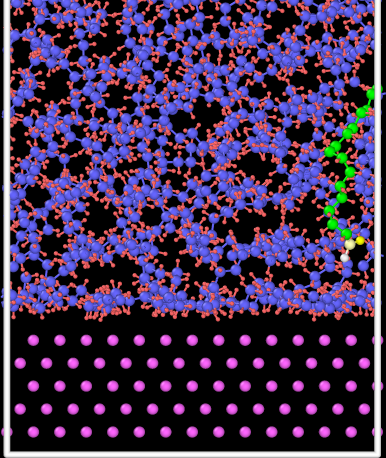}
  \end{center}
  \subcaption{t = 5.50 ns.\\}
\vspace{2cm}
  \label{fig:snapshot1}
 \end{minipage}
 \begin{minipage}{0.5\hsize}
  \begin{center}
   \includegraphics{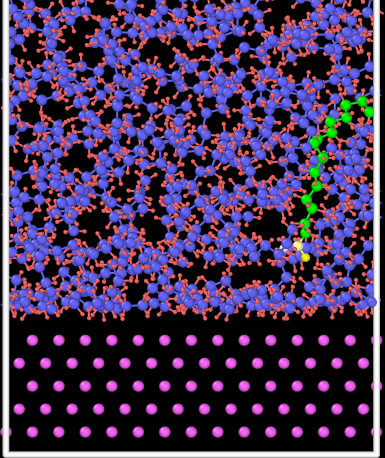}
  \end{center}
  \subcaption{t = 5.52 ns.\\}
\vspace{2cm}
  \label{fig:snaphot2}
 \end{minipage}\\
  \begin{minipage}{0.5\hsize}
  \begin{center}
    \includegraphics{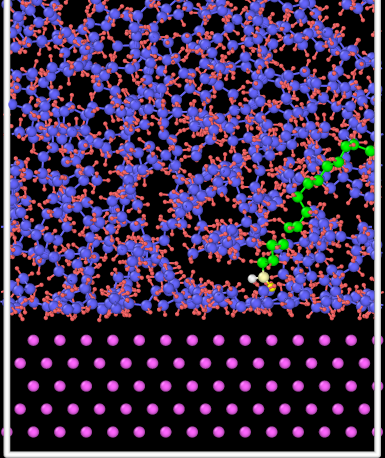}
  \end{center}
  \subcaption{t = 5.54 ns.}
  \label{fig:snapshot3}
 \end{minipage}
   \begin{minipage}{0.5\hsize}
  \begin{center}
    \includegraphics{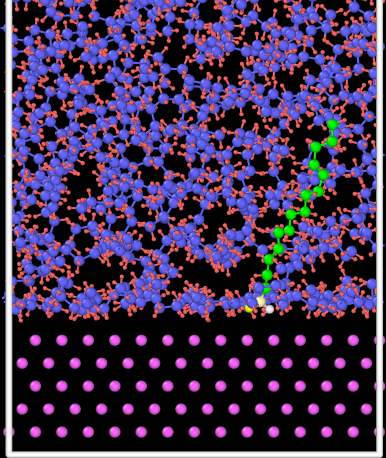}
  \end{center}
  \subcaption{t = 5.56 ns.}
  \label{fig:snapshot4}
 \end{minipage}
   \caption{Snapshots of the adsorption process of
     palmitic acid molecule in n-hexadecane solution
     to the charged surface.
     The Fe solid atoms are charged in t~=~0 ns.
     The colors of the atoms are the same as in
   Fig.\ref{fig:model_make}. }
 \label{fig:snapshots}
\end{figure}

Figure~\ref{fig:snapshots} shows snapshots of the adsorption process of the palmitic acid
molecule in the n-hexadecane solution on the charged surface. 
It is obvious that the base oil molecules construct the absorbed layer in the
vicinity of the solid wall. 
The phenomena are suggested by many
studies~\cite{washizu_lubri, Washizu:2014, Israelachvili:2010, Abraham:1978, Bistans:1987}.
The fluid molecules show an oscillating density profile at the interface 
of the solid atoms~\cite{Abraham:1978}
due to the cohesive force arising from the attraction potential of the solid atoms to the fluids.
In this solution system, since the number of base oil atoms is much higher than the solute palmitic acid,
the adsorbed layer of the base oils behave as the wall to inhibit the adsorption of the
solute molecules.
Subsequently, the palmitic acid molecule breaks through the absorbed layer of the base oils
at t~=~5.56~ns 
(this time depends on the initial condition)
shown in Fig.~\ref{fig:snapshots} (d).
This is due to the long-range Coulomb force between the solid atoms and the carboxyl group of the
palmitic acid.

\begin{figure}[htbp]
 \begin{minipage}{0.2\hsize}
  \begin{center}
    \includegraphics{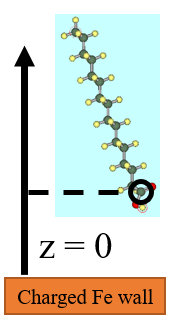}
  \end{center}
  \subcaption{Schematic picture of the coordinate and
    the position of the carbon atom which are sampled.
  }
  \label{fig:plot_point}
 \end{minipage}
 \begin{minipage}{0.35\hsize}
  \begin{center}
   \includegraphics{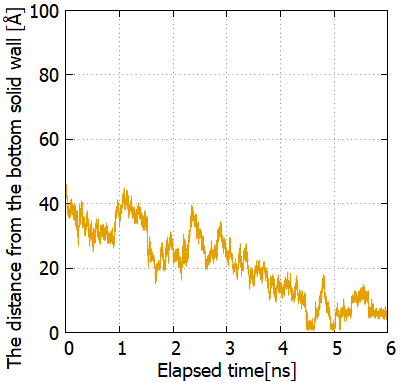}
  \end{center}
  \subcaption{The trajectory in the $z$ direction of the adsorbed molecule.}
  \label{fig:snaphot2}
 \end{minipage}
  \begin{minipage}{0.35\hsize}
  \begin{center}
    \includegraphics{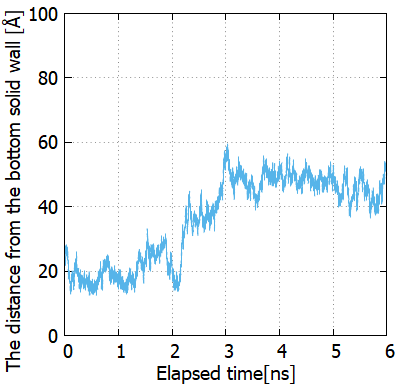}
  \end{center}
  \subcaption{The trajectory in th $z$ direction of the molecule which is not adsorbed.}
  \label{fig:snapshot3}
 \end{minipage}
  \caption{Time evolution of the position of the additive molecule
    in direction $z$ perpendicular to the solid surface.}
 \label{fig:time_history}
\end{figure}

The strength of the layer formation of the base oil molecules is also shown
by the time development of the fluctuation in the $z$ direction of 
the palmitic acid molecules. 
Figure~\ref{fig:time_history} shows the time history of two palmitic acid molecules
selected from the same ensemble of the solution.
In the time development of the 
finaly adsorbed molecule shown in Fig.~\ref{fig:time_history} (b),
the molecule first stays at $z = 20.0$~\AA, then
stays around this height. After 1.5~ns, it start to descend from
$z = 20.0$~\AA~to $z = 10.0$~\AA~at t~=~4.3 ns and suddenly
touches the solid surface. 
In the time development of the 
finally not adsorbed molecule shown in Fig.~\ref{fig:time_history} (c),
the molecule fluctuates around $z = 20.0$~\AA~for about 2~ns, 
then moves in the opposite direction from the surface. 
This results shows that about a 20~\AA~thick layer inhibits the palmitic
acid molecules to form touching the surface of the solid layer by
forming the adsorbed layer. 

In the initial condition, in all case, 
the nearest molecule to the solid surface position are in $z = 20.0$~\AA.
However, this does not mean the nearest molecule first reach 
to the solid surface. Even if the initial distribution of the
molecules are same, the movement of molecules are very different
due to the difference of the initial velocity.

\begin{figure}[htbp]
 \begin{center}
  \includegraphics{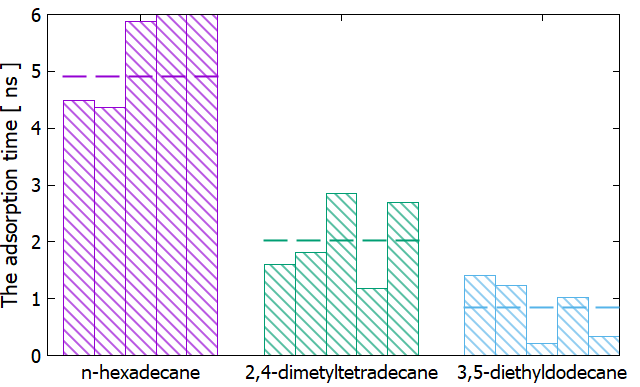}
 \end{center}
 \caption{Adsorption time in each base oil. Each bar shows the
   time which the first additive molecule approached the
   surface of the solid using the same initial configuration and
   different set of initial velocities. The dashed line shows
   the average adsoption time in the 5 trajectories.
   In n-hexadecane, since 2 system does not show an adsorption,
   the average time is calculated from the former 3 trajectories.}
 \label{fig:adsorption_time}
\end{figure}

Figure~\ref{fig:time_history} also shows that the time for the
first molecule to reach the surface takes about 4.5~ns.
The question then arises whether this time will change
due to the structure of the base oil molecule.
Figure~\ref{fig:adsorption_time} shows the adsorption time, 
which is the time 
the palmitic acid molecules first reaches the surface of the
solid layer in the 3 base oils, i.e.,
n-hexadecane and 2, 4-dimethyltetradecane, 
and 3, 5-dimethydodecane.
In each solution, 5 simulations are done by changing
the seed of the random number which decides the initial
velocity of each atom in the solution.
The graph even shows that the difference in the adsorption
time in the same base oil molecules is high and
the adsorption time between the base oil molecules
remarkably differs. 
The branhced molecules show half the adsorption time
vs. the linear molecules, and the time also differs
between the two branched molecules.
If the length of the branched chain is long
(3, 5-dimethydodecane), the adsorption time is long.
Therefore, we confirmed that 
not only the structure~\cite{Dijkstra}, but
the adsoption time due to 
the structure of the base oil molecules can be detected
using the molecular dynamics simulation.

The difference in the adsorption time may due to structuring of
the base oils.
Structuring of the base oil in the systems of a branched alkane
is weaker than that in the system of a linear alkane. 
It is thought that the branched alkane side-chains reduce 
the pair-potential vibration~\cite{Israelachvili:2010}.

\begin{figure}[htbp]
 \begin{center}
  \includegraphics{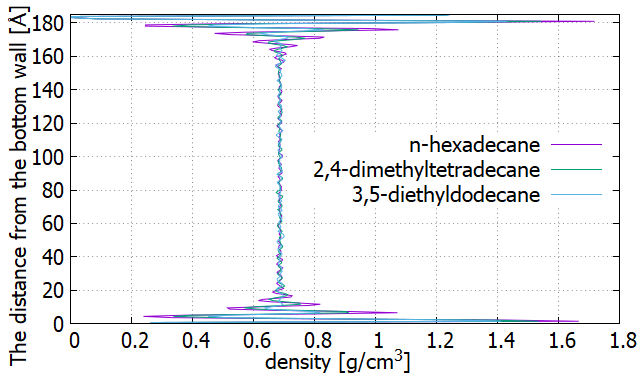}
  \includegraphics{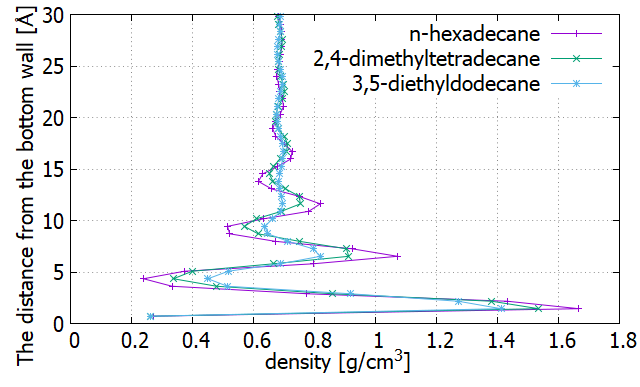}
 \end{center}
 \caption{(a) Profiles of carbon atom density of the three
   base oil molecules
    in direction $z$ perpendicular to the solid surface.
    (b) Enlarged view of (a).
}   
 \label{fig:density_log}
\end{figure}

In order to understand the mechanism, density profiles of each of the base oil 
molecules are plotted in Fig.~\ref{fig:density_log}.
The oscillation of the density is clearly shown in the vicinity of the solid atom,
and the peak differs between the base oil molecules.
The graph also shows that even the system in the $z$ direction is asymmetric
due to the difference between the charged and not-charged walls,
while the distribution the of base oil is not almost different. 
This is because the base oil molecules, which are made of alkyl chains, 
are not polarized, so the charge on the solid layer
does not affect the structuring, and the van der Waals interaction becomes
dominant.
In the center of the oil film, a plateau region of the density profiles
is found. This means that the film thickness of our simulation is large enough 
so that the adsorption process from the bulk region to the structured region
can be discussed~\cite{Doig, Ewen2018}.

In order to see the precise distribution of the base oil molecules,
the enlarged view of Fig.~\ref{fig:density_log} (a) is plotted in
Fig.~\ref{fig:density_log} (b).
It is clearly shown that the order of the 
structuring of the base oil is 
n-hexadecane, 2, 4-dimethyltetradecane, 
and 3, 5-dimethydodecane, which is consistent with
the length of the adsorption time.

\begin{figure}[htbp]
 \begin{center}
  \includegraphics{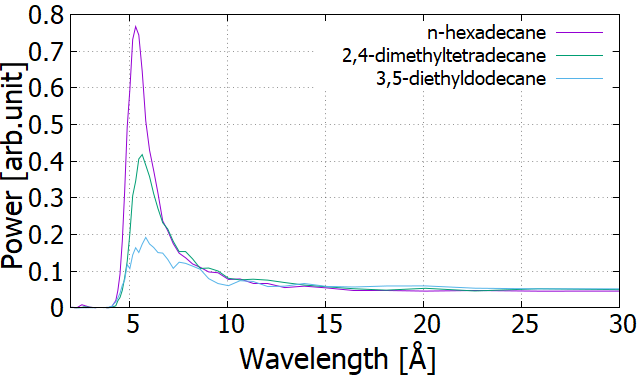}
 \end{center}
 \caption{Fourier transform spectrum of molecular density profiles
   of the three base oil molecules, 
   taken from Fig. \ref{fig:density_log}.
 }   
 \label{fig:denisity_log_fourie}
\end{figure}

The Fourier power spectrum of the density profile
can be used to analyze the strength of the order and periodicity of 
the molecular distribution~\cite{washizu_lubri}.
Figure~\ref{fig:denisity_log_fourie} shows
the Fourier power spectrum obtained from Fig.~\ref{fig:density_log}.
The spectrum is calculated using the Octave FFT analyzer~\cite{Octave}, 
and the number of data points is 512 using the rectanglular
window as the window function.
The peak of the wavelength is about 5 to 6 \AA, which is
equal to the width of the alkyl chain. 
The order of the 
structuring of the base oil is 
n-hexadecane and 2, 4-dimethyltetradecane, 
then 3, 5-dimethydodecane. 
The peak of 3, 5-dimethydodecane is very broad 
compared to the peak of n-hexadecane.
Therefore, we can confirm that the structuring of the base oil
is very different between the structure of the base oil molecules,
and the more structured system takes more time for the additive
molecules to adsorb onto the surface.

\begin{figure}[htbp]
 \begin{center}
  \includegraphics{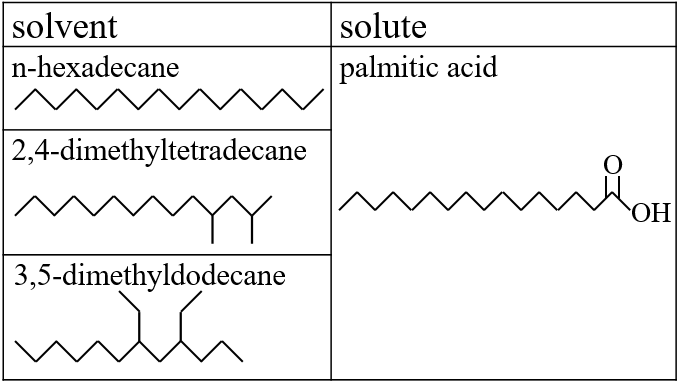}
 \end{center}
  \caption{Molecular structure of 
    base oils (base oils) and solute (oilness additive)
    used in our simulation.}
 \label{figure:table}
\end{figure}

The structure of the base oil at the interface is also
understood by the order parameter. 
The order parameter is defined as the sum of the angle
$\theta$ between the direction of the $xy$ plane and the 
direction of the end-to-end vector of the base oil molecules.
The end-to-end vector is defined as the vector between
the most topologically separated carbon atoms of each
molecules. 
In 2, 4-dimethyltetradecane, one carbon atom at the
end of alkyl chain in the branched side (see Fig.\ref{figure:table})
is randomly chosen as the one side.
The order parameter $P (z)$ is then defined as follows:
\begin{equation}
\label{eq:haikou}
P(z) = ( 3 \langle \cos^2 \theta \rangle - 1) / 2
\end{equation}
where $<>$ denotes the ensemble average and the
$z$ is taken from the center of the mass of each
molecule. 
$P(z)=1.0$ when the end-to-end vector is in the same direction of
the $xy$ plane. When the end-to-end vector is random,
$P(z)$ is zero since $\cos^2 \theta= 1/3$ in the random distribution
of $\theta$. 

\begin{figure}[htbp]
 \begin{center}
  \includegraphics{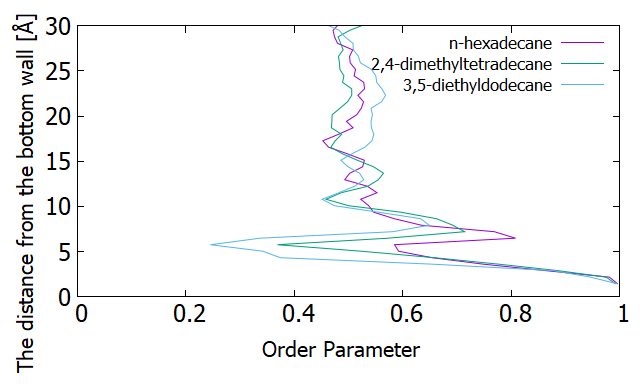}
 \end{center}
 \caption{Distribution of the order parameter of the three base oil molecules
   in the vicinity of the surface.
 }
 \label{fig:angle_order2}
\end{figure}

Figure~\ref{fig:angle_order2} shows the order parameter
$P (z)$ in the vicinity of the solid surface.
In all types of base oil molecules, $P (z)$ is almost 1,
since the center of mass is the lowest $z$ value,
and almost all the carbon atoms are in the lowest $z$ coordinate.
The $P (z)$ suddenly decreases with the increase in $z$ and
shows a minimum when $z$ is about 5.0~\AA.
Comparing this graph with Fig.~\ref{fig:density_log} (b),
this region corresponds to the intermediate part of the
molecular layer. When the center of mass is 
in this region, molecules are bridging between the two
neighbor molecular layers. The $\theta$ then shows
a higher value, which means the molecules are more
against the $xy$ plane.
The $P (z)$ then increases again, since this region is
the second molecular layer counted from the surface.
The $P (z)$ then periodically oscillates as the $z$ increases
and reaches to the plateau value.
Comparing the base oil molecules in this region,
n-hexadecane shows the most ordered structure 
and 2, 4-dimethyltetradecane and 3, 5-dimethydodecane
are less structured. Therefore, we can understand
from this viewpoint that the linear alkane molecules
are much more structured than the branched molecules.
The steric hindrance of the linear alkane molecules
are much larger than the branched molecules.
If the order parameter is small which mean
the molecules are randomly distributed, 
there may be a gap which the additive molecule
can pass through, but when the base oil make a
a more structured layer as in the linear alkane,
such as n-hexadecane, the additive molecule
is hard to go through a small gap.

The adsorption process is a dynamic process in which the additive
molecules diffuse through the base oil layers and
reach the solid surface.
In order to understand the dynamics, 
analyzing the diffusion of the additive molecules is 
the best way to show the adsorption mechanism.
The number of additive molecules, however, is
too low in the MD simulations, thus it is very hard to
directly obtain the diffusion coefficient of the additive
molecules.
Even if the avaraged diffusion constant of additive molecules
are obtained, the critical diffusion process is the
diffusion in the vicinity of the surface. 
Therefore, the distribution of diffusion coefficient is
needed to discuss. In the dilute solution under considering,
the diffusion coefficient of additive molecules shows
large fluctuation.
Therefore, we calculated the diffusion coefficients
of the solvent base oil molecule, in order to
clarify the mechanism.
The diffusion coefficient of the additive moleule and
linear alkane may be similar, since the molecular weight
and structure is similar, and interactino between 
polar groups would not affect in dilute solution.
The diffusion coefficient of the additive moleule and
branched alkane may be different. Therefore, this 
analysis of diffusion may be completed in the later
studies using more powerful computers.

\begin{figure}[htbp]
  \begin{center}
    \includegraphics{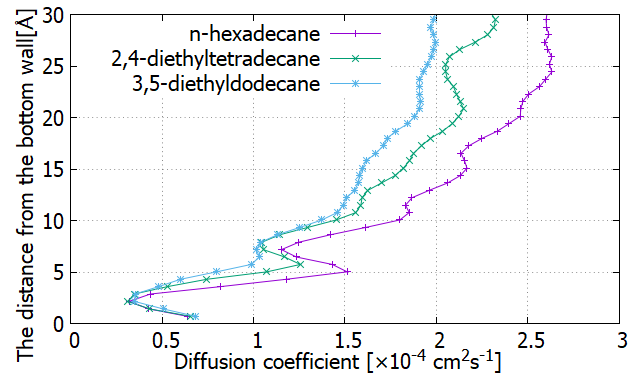}
  \end{center}
 \caption{Diffusion coefficients of the three base oil molecules
   in the vicinity of the surface.
 }
  \label{fig:diff_coeff_zoom}
\end{figure}

Figure~\ref{fig:diff_coeff_zoom} shows the 
self-diffusion coefficients $D_{{s}} (z)$ as the function
of the $z$ coordinate. 
$D_{{s}} (z)$ is calculated by the following equation.
\begin{equation}
\label{eq:einstein}
D_{{s}} (z) = \lim_{t \to \infty} \frac{1}{2 t} \langle | z(t) - z(0) |^2 \rangle,
\end{equation}
where $t$ is the time and $z(t)$ is the $z$ coordinate at time $t$.
$D_{{s}} (z)$ is calculated for every 20~ps of motion of the
base oil molecules so that the long-range dynamics are taken.
In Fig.~\ref{fig:diff_coeff_zoom}, $D_{{s}} (z)$ decreases as the
$z$ coordinate decreases from the center of the oil film.
This is because the structured oil film in the vicinity of the interface
affects not only the static structure but the dynamics is different
in the interface.
The peak shown at the very low $z$ is due to the rebound motion of the
reflective solid layer. Each base oil molecule almost stops the
motion at $z$ = 2.5~\AA which is the peak of the
first molecular layer shown in  Fig.~\ref{fig:density_log}.
$D_{{s}} (z)$ then shows the minimum at the intermediate region
of the layers and shows maximum at the peak of the layers.
Comparing the three base oils, 
the $D_{{s}} (z)$ value of n-hexadecane is the 
highest except in the first layer. 
This means that the motion of the molecules are not
suppressed in the whole solution, but 
the structured molecular layer, especially at the
first layer, acts as the hardest barrier to the 
adsorption process. 

The diffusion coefficients in bulk calculated by
the MD simulation are not well predicted in
some studies~\cite{diff_sim1, diff_sim2, diff_exp}.
However, our calculated diffusion coefficient
is on the same order (1.0$\times 10^{-5} {cm}^2 {s}^{-1}$).

Usually, the adsorption dynamics of molecules are
separated into two process, i.e., the adsorption limited process
and diffusion limited process~\cite{miura}. 
The former is the difficulty of the physical adsorption
using the surface interaction. The latter is the diffusion
of the additive molecule in the base oil.
In a previous study using coarse grained molecular
dynamics~\cite{miura}, 
it showed that the diffusion limited adsorption is
the major process. 
In our study, however,
the motion of the additive molecule to break through
the first structured layer of the base oil is the
limiting process, which mean 
although the ratio of the importance of both 
process can not be determined, since the simulation
time and space is limited,
adsorption limited process is also important.

The diffusion coefficient is rather higher in
the linear alkane but the adsorption time is
longer in it.
The difference in the phenomena
is understood by the difference
in the base oil model we used.

In real phenomena, 
this means that 
not only the structure of additive molecules
~\cite{Loehle2014}, but
the structure of
the base oil is important to evaluate the
efficiency of the surface protection.
The formation time of the adsorbed additive
layer, 
as well as the coverage ratio
~\cite{Ewen2016a}, 
is important 
in the case such as
the sliding speed of the
machine element is high
and the pressure is high enough to break
the adsorbed layer.
If the sliding condition is mild,
the stability of the adsorbed layer
is more important, which mean the
chain matching phenomena may be dominant.
There are few studies which take into
consideration the formation process
of the adsorbed additive layer in tribology.
We would like to note that structuring
of the base oil on the surface is very
important for the friction control
using OFMs. 

A further study may include the succeeding
process of adsorbed additive layers, 
concentration dependence of
additive molecules, and
the chemical effect of surface and additives.

\section{Conclusions}

Using an all-atom molecular dynamics with an explicit base oil, 
the phenomena that the molecules of the additives break through 
the adsorbed layer of the base oil is observed.
The relative adsorption time depends on the structure of the base oil molecules. 
The adsorption process is hence not a diffusion limited but adsorption limited process.
The result reveals that the structuring of the base oil molecules 
near the solid wall causes a limitation for the adsorbing process.
This phenomena will result in the boundary lubrication effect of the machine elements.

\section*{Acknowledgements}

This study was supported by JSPS (Japan Society for the Promotion of Science) KAKENHI (Grants-in-Aid for Scientific Research) Challenging Research (Exploratory) Grant Number 18K18813.
We also thank Dr. Hiroaki Koshima and Dr. Kazuhiro Yagishita for
their usefull discussions.

\newpage

\newpage

\newpage

\newpage

\vspace{5cm}

\section*{References}

\begin{thebibliography}{30}
\bibitem[Bowden and Tabor(1964)]{Bowden:1964}
F.~Bowden and D.~Tabor, in \emph{The friction and lubrication of solids},
  Oxford: Clarendon Press, Oxford, 1964\relax

\bibitem{Spikes:2015}
  H. Spikes, 
  Tribol Lett, 60:5 (2015).
  
\bibitem{Askwith:1966}
  T. C. Askwith, A. Cameron, R. F. Crouch,
  Proc. R. Soc. Lond. A, 291, 500-519, (1966).
  
\bibitem{Hirano:1987}
  F. Hirano, T. Sakai, K. Kuwano, N. Ohno,
  Trib. Intl., 20, 4, 186 (1987).
  
\bibitem{Cameron:1997}
  A. Cameron,
  Tirb. Lett. 3, 13 (1997).
  
\bibitem{koike}
A. Koike, and M. Yoneya, Molecular dynamics simulations of sliding friction of Langmuir-Blodgett monolayers, J. Chem. Phys, 105, 6060, (1996).

\bibitem{koike2}
A. Koike, and M. Yoneya, Effects of Molecular Structure on Frictional Properties of Langmuir-Blodgett Monolayers, Langmuir 13, 1718, (1997).

\bibitem{ohzono}
T. Ohzono, M. Fujihira, Molecular dynamics simulations of friction between an ordered organic monolayer and a rigid slider with an atomic-scale protuberance, Phys. Rev. B, 62, 17, 055-071, (2000).

\bibitem{cummings}
Y. Leng, D. J. Keffer, and P. T. Cummings J. Phys. Chem. B, 107, 43, 11940, (2003).

\bibitem{Doig}
M. Doig et al. Structure and Friction of Stearic Acid and Oleic Acid Films Adsorbed on Iron Oxide Surfaces in Squalane. Langmuir 30, 186-195 (2014).

\bibitem{Ewen2016a}
J. P. Ewen et al. Nonequilibrium Molecular Dynamics Simulations of Organic Friction Modifiers Adsorbed on Iron Oxide Surfaces. Langmuir 32, 4450-4463 (2016).

\bibitem{Ewen2018}
J. P. Ewen et al. Slip of Alkanes Confined between Surfactant Monolayers Adsorbed on Solid Surfaces. Langmuir 34, 3864-3873 (2018).

\bibitem{Bradley-Shaw2016}
J. L. Bradley-Shaw et al. Molecular Dynamics Simulations of Glycerol Monooleate Confined between Mica Surfaces. Langmuir 32, 7707-7718 (2016).

\bibitem{Bradley-Shaw2018}
J. L. Bradley-Shaw et al. Self-assembly and friction of glycerol monooleate and its hydrolysis products in bulk and confined non-aqueous solvents. Phys. Chem. Chem. Phys. 20, 17648-17657 (2018).

\bibitem{Loehle2014}
S. Loehle et al. Mixed Lubrication with C18 Fatty Acids: Effect of Unsaturation. Tribol. Lett. 53, 319-328 (2014).

\bibitem{Loehle2015}
S. Loehle et al. Mixed lubrication of steel by C18 fatty acids revisited. Part I: Toward the formation of carboxylate. Tribol. Int. 82, 218-227 (2015).

\bibitem{Gattinoni}
C. Gattinoni et al. Adsorption of Surfactants on <alpha>-Fe2O3(0001): A Density Functional Theory Study. J. Phys. Chem. C 122, 20817-20826 (2018).

\bibitem{Campana2011}
M. Campana et al. Surfactant adsorption at the metal-oil interface. Langmuir 27, 6085-6090 (2011).

\bibitem{Campana2015}
M. Campen et al. In Situ Study of Model Organic Friction Modifiers Using Liquid Cell AFM; Saturated and Mono-unsaturated Carboxylic Acids. Tribol. Lett. 57, 18 (2015).

\bibitem{Ewen2016b}
J. P. Ewen et al. A Comparison of Classical Force-Fields for Molecular Dynamics Simulations of Lubricants. Materials 9, 651 (2016).

\bibitem{miura}
T. Miura, M.Mikami, Molecular dynamics study of the effects of chain properties on the order formation dynamics of self-assembled monolayers of long-chain molecules, Phys. Rev, E 81, 021801, (2010).

\bibitem{dreiding}
S. L. Mayo, B. D. Olafson, DREIDING:A Generic Force Field for Molecular Simulations. J.Phys. Chem. 94, 8897-8909, (1990).

\bibitem{mopac}
Computational Chemistry, David Young, Wiley-Interscience, 2001. Appendix A.A.3.2 pg 342, MOPAC.

\bibitem{chainmatching_ori}
T. C. Askwith, A. Cameron, and R. F. Crouch, Proc. Roy. Soc. London, A 291, 500, (1966).

\bibitem{lammps}
S. Plimpton, Fast Parallel Algorithms for Short-Range Molecular Dynamics, J. Comput. Phys., 117, 1 (1995).

\bibitem{nvt}
S.Nos\'{e}, A unified formulation of the constant-temperature molecular dynamics methods, J.Chem.Phys. 81, 511, (1984) .

\bibitem{hoover}
W. G. Hoover, Canonical dynamics: Equilibrium phase-space distributions, Phys. Rev. A,31, 1695, (1986).

\bibitem{nvt_text_ref}
S.Nos\'{e}, A Molcular Dynamics Method for Simulations in the Canonical Ensemble, Molec.Phys., 52, 255, (1984).

\bibitem{npt}
H.C. Anderson, Molecular Dynamics Simulations at Constant Pressure and/or Temperature, J. Chem. Phys., 72, 2384, (1980).

\bibitem{msm}
D. J. Hardy, J. E. Stone, K. Schulten, Parallel Comput. 35, 164 (2009).

\bibitem{verlet}
L. Verlet, Computer "Experiments" on Classical Fluids. I. Thermodynamical Properties of Lennard-Jones Molecules, Phy.Rev. 159, 98, (1967).

\bibitem{Washizu:2018}
K. Nishikawa, H. Akiyama, K. Yagishita, H. Washizu, "Molecular dynamics analysis of adsorption process of anti-copper-corrosion additives to the copper surface", arXiv:1812.10647 (in printing).

\bibitem{washizu_lubri}
H. Washizu, T. Ohmori. Molecular Dynamics Simulations of Elastohydrodynamic Lubrication Oil Film, Lubrication Sciences, 22, 323, (2010)．

\bibitem{Washizu:2014}
H. Washizu, S. Hyodo, S. Ohmori, N. Nishino, A. Suzuki, Macroscopic no-slip boundary condition confirmed in full atomistic simulation of oil film, Tribology Online 9, (2) 45-50 (2014).

\bibitem{Washizu:2017}
H. Washizu, T. Ohmori, A. Suzuki, "Molecular Origin of Limiting Shear Stress of Elastohydrodynamic Lubrication Oil Film Studied by Molecular Dynamics", Chem. Phys. Lett.,678, 1-4 (2017). 

\bibitem{Tamura}
H. Tamura et al., Molecular Dynamics Simulation of Friction of Hydrocarbon Thin Films,
Langmuir, 15 (22), 7816–7821 (1999).

\bibitem[Israelachvili(2010)]{Israelachvili:2010}
Israelachvili, in \emph{Intermolecular and Surface Forces, Third Edition},
  Academic Press;, Burlington, 2010\relax

\bibitem{Abraham:1978}
F.F. Abraham, J. Chem. Phys. 68, 3713 (1978).
 
\bibitem{Bistans:1987}
I. Bitsanis et al., J. Chem. Phys. 87, 1733 (1987).

\bibitem{Dijkstra}
M. Dijkstra, J. Chem. Phys. Confined thin films of linear and branched alkanes. 107, 3277-3288 (1997).

\bibitem{Octave}
J. W. Eaton, D. Bateman, S. Hauberg, R.Wehbring,
{GNU Octave} version 3.8.1 manual: a high-level interactive language for numerical computations,
CreateSpace Independent Publishing Platform (2014).

\bibitem{diff_sim1}
E. B. Webb I\hspace{-.1em}I\hspace{-.1em}I,  Gary S. Grest, and Maurio Mondello.  Intracrystalline Diffusion of Linear and Branched Alkanes in the Zeolites TON, EUO, and MFI,  J. Phys. Chem. B,103 ,4949-4959, (1999).

\bibitem{diff_sim2}
K. S. Kostov et al., Dynamics of linear and branched alkane melts: Molecular dynamics test of theory for long time dynamics, J.Chem. Phys. 108, 21, (1998). 

\bibitem{diff_exp}
D. Alonso De Mezquia et al., Determination of Molecular Diffusion Coefficient in n-Alkane Binary Mixtures: Empirical Correlations, J. Phys. Chem. B, 116, 2814-2819, (2012). 

\end{thebibliography}
\end{document}